\documentclass[adp,a4paper,fleqn]{w-art}
\usepackage{times,cite,w-thm}
\usepackage[pdftex]{graphicx}
 
\begin{document}

\DOIsuffix{theDOIsuffix}
\Volume{16}
\Month{01}
\Year{2007}
\Receiveddate{XXXX}
\Reviseddate{XXXX}
\Accepteddate{XXXX}
\Dateposted{XXXX}

\title{The Quantum Boltzmann Equation in Semiconductor Physics}
\author{D.W. Snoke}
\address{Department of Physics and Astronomy, University of Pittsburgh, Pittsburgh, PA, 15260, USA}

\begin{abstract}
The quantum Boltzmann equation, or Fokker-Planck equation, has been used to successfully explain a number of experiments in semiconductor optics in the past two decades. This paper reviews some of the developments of this work, including models of excitons in bulk materials, electron-hole plasmas, and polariton gases. 
\end{abstract}

\maketitle

\section{Introduction}

Already when Boltzmann was working on the foundations of statistical mechanics in the late 1800's, it was widely recognized that the equilibrium theorems of statistical mechanics relied on an ``H-theorem,'' that is, a theorem that shows that systems out of equilibrium evolve toward equilibrium. Boltzmann's approach to the H-theorem relied on the statistics of classical particles. Since the development of quantum mechanics, in particular the second-quantized field theory formalized by Dirac in the 1930's, it has been understood that an H-theorem can be written down using the time-dependent 
Schr\"odinger equation for many-particle systems. The equation for the time evolution of the distribution function of a quantum many-body system is variously called a quantum Boltzmann equation (after the collision term in the classical Boltzmann equation, discussed in Ref.~\cite{snokebook}, Section 5.9), a master equation (when the states are discrete), or a Fokker-Planck equation. 

While the formalism of time-dependent quantum mechanics of many-particle systems was well developed in the mid-twentieth century, two developments occurred in the 1980's that allowed much more quantitative application to experiments on nonequilibrium systems. The first was the arrival of cheap, fast computers that allowed numerical solution of the quantum Boltzmann equation using iterative methods. The second was the development of ultrafast optics methods which allowed direct observation of the distribution function of carriers in semiconductors (and in some metals)
out of equilibrium. A related development, in the 1990's, was the accomplishment of trapped cold atom gases, which also allowed direct measurement of the distribution function of a gas. The results of quantum Boltzmann equations could thus be directly compared to experimental measurements of particle distribution functions instead of only average properties. 

In the limit of low particle density, the quantum Boltzmann equation has the form of simply adding all the rates determined by Fermi's golden rule for all possible scattering processes. For evolution due to single-particle transitions, this is given by
\begin{equation}
\frac{\partial n_i}{\partial t} = \sum_j \frac{2\pi}{\hbar}\left|\langle i | H_{\rm int} | j\rangle \right|^2\delta(E_j-E_i)n_j  -\sum_j \frac{2\pi}{\hbar}\left|\langle j | H_{\rm int} | i\rangle\right|^2\delta(E_i-E_j)n_i,
\label{lowd}
\end{equation}
where $n_i$ and $n_j$ are the instantaneous occupation probabilities of states $i$ and $j$. This can be solved numerically by an iterative method in which an instantaneous distribution function $n_i(t)$ is used to calculate all possible rates $\partial n_i/\partial t$, and then the distribution function is updated for all $i$ in the next time step of length $dt$ according to $n_i(d+dt) = n_i(t) + (\partial n_i/\partial t)dt$. 

While this approach makes intuitive sense, justifying the quantum Boltzmann equation quantum mechanically requires a fair amount of effort. The problem is that properly speaking, the time evolution should be applied not to each state individually, but to the full many-body state of all the particles. Equation (\ref{lowd}) is incorrect when the particle density is such that the quantum Bose or Fermi statistics of the particles become important, which corresponds approximately to the density at which the average deBroglie wavelength of the particles is comparable to the average distance between them. It also does not apply when there is phase coherence between the particles, as is the case when there is a Bose condensate, although as discussed below, there are methods to modify the quantum Boltzmann approach to model condensates. Even in the low-density limit, this equation must be modified if there are two-body scattering processes or higher-order scattering processes. 

In general, it is not obvious that using Fermi's golden rule, which is derived for a {\em single} particle, will work in a {\em many}-particle system, where there is the possibility for multiple phase interferences. It turns out that it usually does, because the phases of the various particles all average out in most cases. Section 4.8 of Ref.~\cite{snokebook} gives the general derivation of the quantum Boltzmann equation from first principles. One generally useful example is the case of a two-body scattering process with identical particles, with interaction Hamiltonian
\begin{equation}
H_{\rm int} = \frac{1}{2}\sum_{k_1,k_2,k_3} U(k_1,k_2,k_3,k_4) a^\dagger_{k_4}a^\dagger_{k_3}a^{ }_{k_2}a^{ }_{k_1},
\end{equation}
where $U(k_1,k_2,k_3,k_4)$ is the interaction potential for two incoming momenta $k_1$ and $k_2$ and two outgoing momenta, $k_3$ and $k_4$, with the value of $\vec{k}_4$ determined by momentum conservation, $\vec{k_1} + \vec{k}_2 = \vec{k}_3 + \vec{k}_4$.
The $a$'s and $a^{\dagger}$'s are destruction and creation operators for the particles in states $\vec{k}_1$, $\vec{k}_2$, $\vec{k}_3$, and $\vec{k}_4$. In this case the quantum Boltzmann time evolution equation is found to be
\begin{eqnarray}
\frac{\partial n_k}{\partial t} &=& \frac{2\pi}{\hbar} \sum_{k_1,k_2} |U_D\pm U_E|^2 \delta(E_{k_1}+E_{k_2} - E_{k'}-E_k) n_{k_1}n_{k_2}(1\pm n_{k'})(1\pm n_k)\nonumber\\
&&- \frac{2\pi}{\hbar} \sum_{k_1,k_2} |U_D\pm U_E|^2 \delta(E_{k'}+E_{k} - E_{k_1}-E_{k_2})n_{k}n_{k'}(1\pm n_{k_1})(1\pm n_{k_2}),\label{twobody}
\end{eqnarray}
where $\vec{k}'$ is the determined by momentum conservation, $\vec{k}_1 + \vec{k}_2 = \vec{k}+\vec{k}'$.
The rate of each transition is multiplied by the factor $(1\pm n_f)$ for the number of particles in each final state $f$; since this equation is for a two-body scattering process, there are two such final states factors. The plus sign is used in the above for bosons, and the minus sign for fermions. Thus, fermion scattering is suppressed by high occupation, while boson scattering is enhanced by it.  In Equation (\ref{twobody}), $U_D$ stands for the direct interaction $U(k_1,k_2,k',k)$, while $U_E$ stands for the exchange process, that is, with the final two states exchanged.  Fermion scattering is therefore forbidden if $U(k_1,k_2,k',k)$ is symmetric under exchange, for example if it is constant, as is the case for $s$-wave, pointlike scattering. 

Similarly, the time evolution for particles emitting or absorbing phonons (which are not number conserved) via the interaction Hamiltonian (see Section 5.1 of Ref.~\cite{snokebook}),
\begin{equation}
H_{\rm int} = i\sum_{q,k} U(q) (a_q a^{\dagger}_{k+q}a_k -  a^{\dagger}_q a^{\dagger}_{k+q}a_k),
\end{equation}
is found to be
\begin{eqnarray}
\frac{\partial n_k}{\partial t}  & = & 
\frac{2\pi}{\hbar} \sum_q  |U({\bf q})|^2 
 \delta(E_{{\bf k}+{\bf q}} - E_{\bf k} -\hbar
\omega_{\bf q})  n_{k + q}(1\pm n_k)(1 + n^{\rm ph}_q) \nonumber\\
&& + \frac{2\pi}{\hbar} \sum_q  |U({\bf q})|^2
  \delta(E_{{\bf k}+{\bf q}} +\hbar
\omega_{\bf q} - E_{\bf k} ) n_{k + q}(1+n_k) n^{\rm ph}_q \nonumber\\
&& - \frac{2\pi}{\hbar} \sum_q  |U({\bf q})|^2
  \delta(E_{{\bf k}} - E_{{\bf k}-{\bf q}} -\hbar
\omega_{\bf q}) n_{k}(1 \pm n_{k-q})(1+n^{\rm ph}_q)  \nonumber\\
&& - \frac{2\pi}{\hbar} \sum_q  |U({\bf q})|^2
  \delta(E_{{\bf k}} +\hbar
\omega_{\bf q}- E_{{\bf k}-{\bf q}} ) n_{k}(1+n_{k-q})   n^{\rm ph}_q,
\label{phonon}
\end{eqnarray}
where $n^{\rm ph}_q$ is the phonon particle distribution. The first and third terms correspond to phonon emission, which is stimulated by a $(1+n^{\rm ph}_f)$ factor, since the phonons are bosons, while the second and fourth terms correspond to phonon absorption. 

Although Equations (\ref{twobody}) and (\ref{phonon}) require some detailed calculations to justify, their form is actually quite simple: they correspond to the results of adding up Fermi's golden rule for each possible transition, and multiplying by the final states factor $(1\pm n_f)$ for the occupation of each final particle state affected by a transition. 

\begin{figure}[<float>]
\sidecaption
\includegraphics[width=0.45\textwidth]{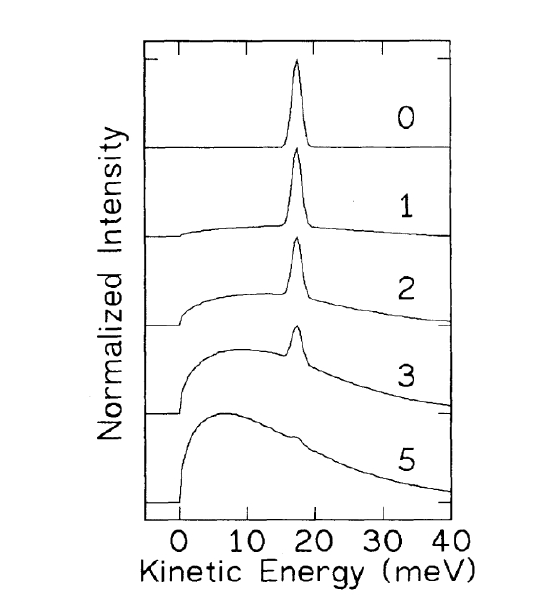}
\caption{Evolution of the distribution function of a homogeneous gas of particles starting far from equilibrum. The labels of the curves are the average number of scattering events per particle, for hard-sphere, two-body scattering. From Ref.~\protect\cite{gaas}.}
\label{fig1}
\end{figure}

Just as classical mechanics is reversible at the microscopic level, so is quantum mechanics. Iterative solution of Equations (\ref{twobody}) and (\ref{phonon}) leads to irreversible behavior, however.  Figure 1 shows the solution of Eq.~(\ref{twobody}) for the distribution function of a homogeneous gas of interacting particles starting far from equilibrium, at low density \cite{snokewolfe,gaas}. (At low density, the evolution is the same for bosons and fermions). The calculated distribution evolves irreversibly toward the equilibrium Maxwell-Boltzmann distribution.
In these calculations, the distribution function in energy, $n(E)D(E)dE = \displaystyle \sum_i n_i\delta(E-E_i)dE$, was evolved directly by the iteration method described above; there was no tracking of individual particles and no random number generation.  It turns out that in the case of a homogeneous, isotropic gas with $U_E$ either negligible or constant, the integrals in (\ref{twobody}) and (\ref{phonon}) can be greatly simplified by integrating over all the angles, leaving only integrals over the energies of the particles. This greatly simplifies the numerical calculations, making calculations like that shown in Fig.~\ref{fig1} possible in just a few hours on a modern personal computer. 

The irreversibility in these calculations comes from the loss of phase information. At each time step in the iteration, the state of the system is represented by a set $\{\ldots, n_{i-1}, n_i, n_{i+1}, \ldots \}$, where the $n_i$ values are continuous variables that give the  occupation probability of each state. This is not the full amount of quantum mechanical information of the system, however. In principle, the full state of the system corresponds to a superposition of all possible states, written as a sum of Fock many-body states, each multiplied by some phase factor $\alpha(t)$ which is a continuous, complex number:
\begin{eqnarray}
|\psi_t\rangle &=& \alpha(t) | \ldots, N_{i-1}, N_i, N_{i+1}, \ldots\rangle + \alpha'(t) | \ldots, N_{i-1}', N_i', N'_{i+1}, \ldots\rangle \nonumber\\
&& +\alpha''(t) | \ldots, N''_{i-1}, N''_i, N''_{i+1}, \ldots\rangle + \ldots,
\label{full}
\end{eqnarray}
where the $N_i$ values are the exact integer values for each state.  The values of $n_i$ correspond to the expectation values of this full quantum mechanical many-body state:
\begin{equation}
n_i = \langle \psi_t | a_i^\dagger a_i | \psi_t\rangle. 
\end{equation}
The quantum Boltzmann equation approach amounts to keeping only the information of these expectation values and not all the individual phase factors.  There is thus information loss at each step, which gives the irreversibility. This is physically well justified in most cases, because the phase factors generally all cancel out for most expectation values except the average occupation numbers.  Exceptions are when the system is very small, and when there is a macroscopic phase, as in the case of a Bose condensate. 

The iterative solution of the quantum Boltzmann equation therefore acts as the H-theorem for any given system, showing that it will evolve toward equilibrium and stay there. The irreversibility in this case comes from dephasing of many-body wave function. This is a generic result in quantum mechanics, that dissipation comes from coupling to many degrees of freedom, which spread out the energy widely.  In principle, one could pick an exact set of phase factors for a full wave function of the form (\ref{full}) which would evolve the system away from an equilibrium probability distribution for a period of time, but for a system which is coupled to an external environment with multiple interactions from other sources, it would be effectively impossible to arrange the phases this way in a system with many particles. 

\section{Results for Excitons} 

\begin{figure}
\includegraphics[width=0.45\textwidth]{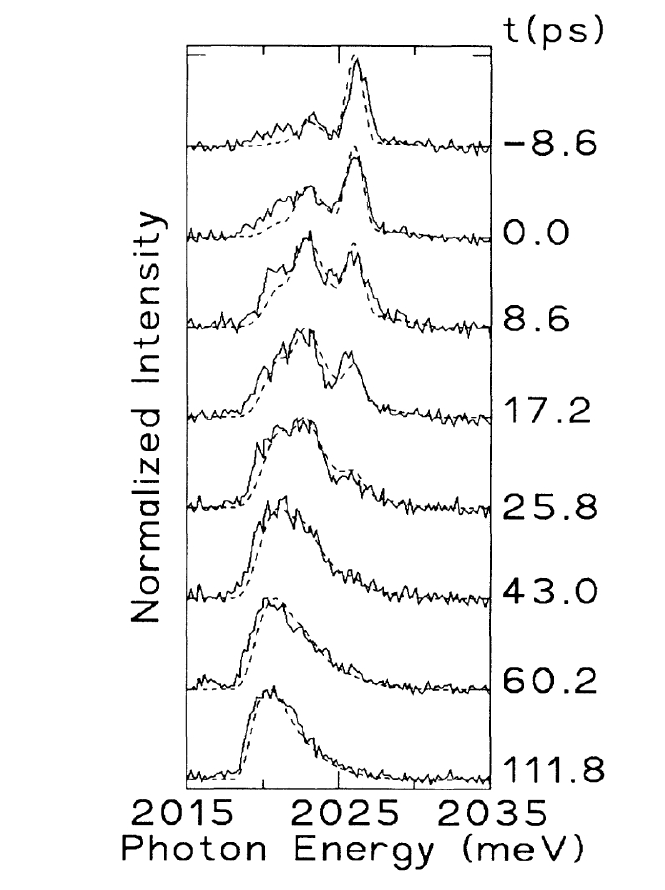}
\includegraphics[width=0.425\textwidth]{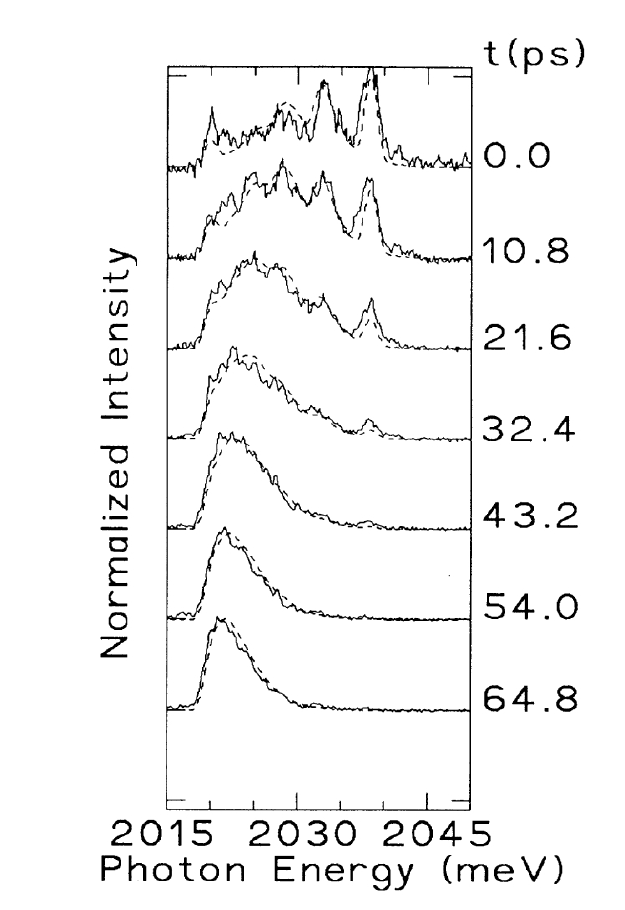}
\caption{Solid lines: experimentally measured evolution of the distribution function of a  gas of excitons in Cu$_2$O starting far from equilibrum, generated by an ultrafast (2 ps) laser pulse at $t=0$, for two different laser photon energies. Dashed lines: theoretical energy distribution at the same times as deduced by iterating the quantum Boltzmann equation. The only fit parameters were the exciton-phonon deformation potentials, which were found to be in agreement with values deduced from stress measurements. From Ref. \protect\cite{snokebc}.}
\label{cu2o}
\end{figure}

As mentioned above, both the power of modern numerical methods as well as the techniques of modern optics allow for a direct comparison of the quantum Boltzmann equation theory to measurements of particle distribution functions. Figure~\ref{cu2o} shows a result by Snoke, Braun, and Cardona \cite{snokebc}, done in the laboratory of Wolfgang Ruehle in Stuttgart in the early 1990's, which compares the evolution of the energy distribution of excitons in the semiconductor Cu$_2$O to the quantum Boltzmann theory, using Equation (\ref{phonon}) with exciton-phonon deformation potentials consistent with static stress experiments in the literature \cite{defpot1,defpot2}. As seen in these data, at early time the exciton distribution evolves in a highly nonequilibrium path, emitting a cascade of phonons, until at late times they reach a Maxwell-Boltzmann distribution. At all later times the distribution remains a Maxwell-Boltzmann, as the temperature cools toward the lattice temperature, which is around 2 K. 

This experiment was possibly the first direct comparison of time-resolved data for a full particle distribution function in any system to a solution of the quantum Boltzmann equation, and remains one of the clearest examples showing that the H-theorem is well justified not only theoretically but also experimentally, as the quantum Boltzmann equation accurately predicts the evolution of the system toward equilibrium. 

This experiment also was significant in that it addressed a controversy about the difference between multi-phonon resonant Raman scattering and hot luminescence. As seen in Fig.~\ref{cu2o}, at early times there are a series of peaks corresponding to emission of various phonons; these peaks last much longer than the laser pulse, although much less than the total exciton lifetime. In a time-integrated spectrum, these peaks will appear on top of a Maxwell-Boltzmann distribution.  Early experiments \cite{yu} interpreted these peaks as multi-phonon resonant Raman lines, but this experiment shows that the most natural interpretation of these lines is as hot luminescence caused by the dominance of certain emission energies due to momentum and energy conservation in the exciton-phonon interaction. The interpretation of these peaks as multi-phonon resonant Raman lines is still widely quoted in the literature \cite{yucardbook}. In principle, all luminescence processes following optical excitation can be seen as a type of Raman process, but when there is a time delay for the emission, following a real absorption process, the picture of hot luminescence is more natural. 

\begin{figure}[<float>]
\sidecaption
\includegraphics[width=0.47\textwidth]{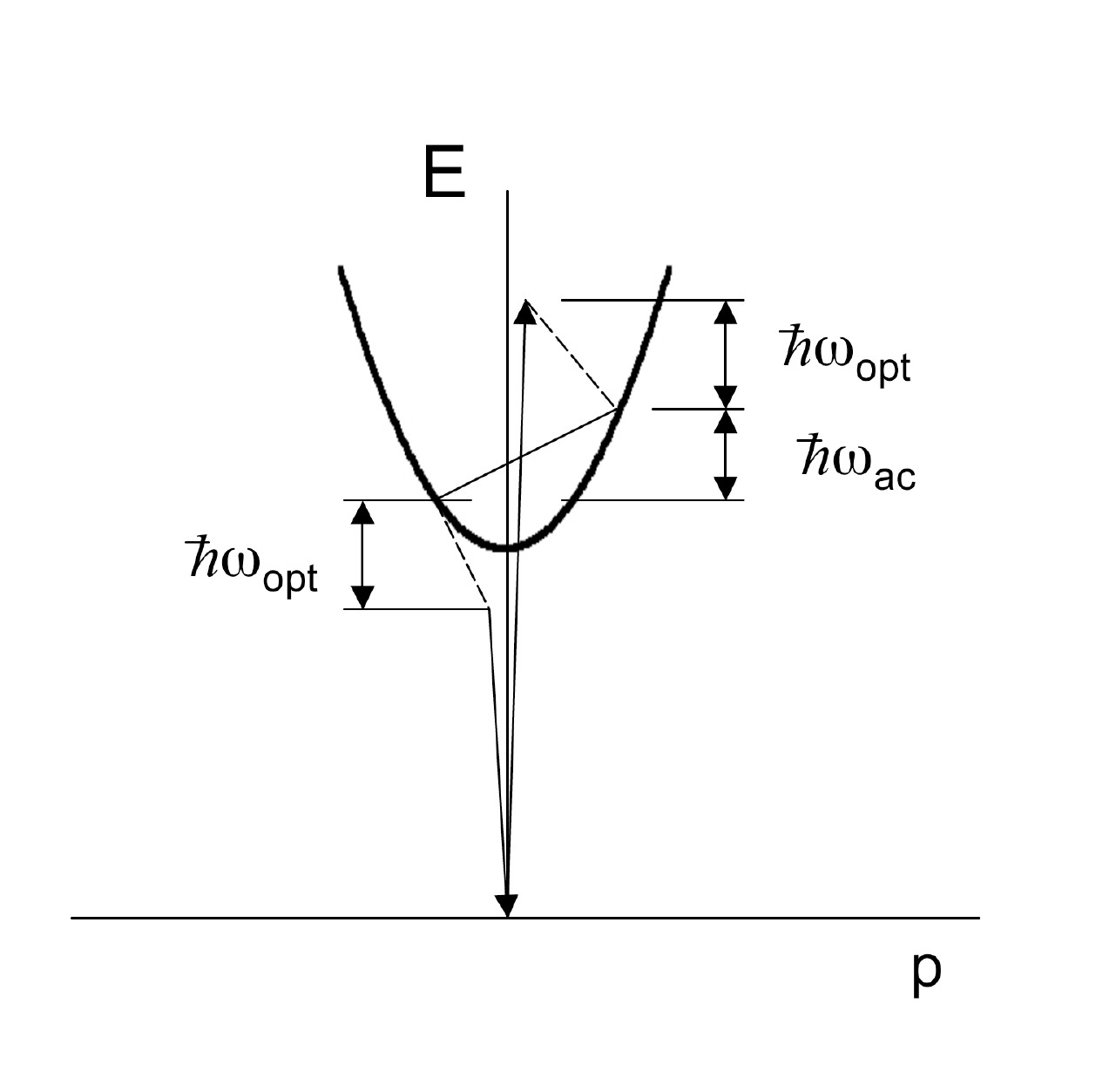}
\caption{Photon absorption and re-emission process in the semiconductor Cu$_2$O. The exciton energy is shown as the solid black parabola. Optical phonon emission is indicated by dashed lines.}
\label{cu2oproc}
\end{figure}

One reason for the debate is the importance of the phonon-assisted exciton absorption and emission in Cu$_2$O. Figure \ref{cu2oproc} illustrates the sequence of processes involved. First, a photon is absorbed along with emission of an optical phonon.  The exciton created by this process can then emit an acoustic phonon. Energy and momentum conservation of this acoustic phonon emission process lead to a peak in the exciton occupation at an energy below the initial exciton energy that depends on the mass of the exciton and the acoustic phonon velocity. The excitons at this energy can then recombine into photons that are emitted along with another optical phonon. Thus, in the final emission spectrum there will be a peak that is two optical phonon energies
below the initial laser energy, due to the optical phonons emitted in the phonon-assisted absorption and emission processes. This will follow the laser if the laser wavelength is tuned. In addition, there will be another peak which is two optical phonons and one acoustic phonon energy lower, and, as seen in Fig.~\ref{cu2o}, there can even be more peaks at lower energy due to additional acoustic phonon emissions. All of these peaks are due to luminescence from the excitons following their creation by the laser. The Boltzmann equation fit shows that all the peaks can be accounted for by the emission and absorption of phonons by hot excitons without need for invoking multi-phonon resonant Raman theory.

In principle, the excitons will scatter with each other and not only with phonons, if the exciton density is high enough. This exciton-exciton scattering would be seen in a faster equilibration of the excitons at higher density, and would therefore constitute a measurement of the exciton-exciton interaction cross section. In the experiments shown in Fig.~\ref{cu2o}, however, it was not possible to have exciton density high enough to see this, due to the long absorption length of the laser near the exciton resonance.

\section{Results for Electron-Hole Plasma}

\begin{figure}
\begin{center}
\includegraphics[width=0.63\textwidth]{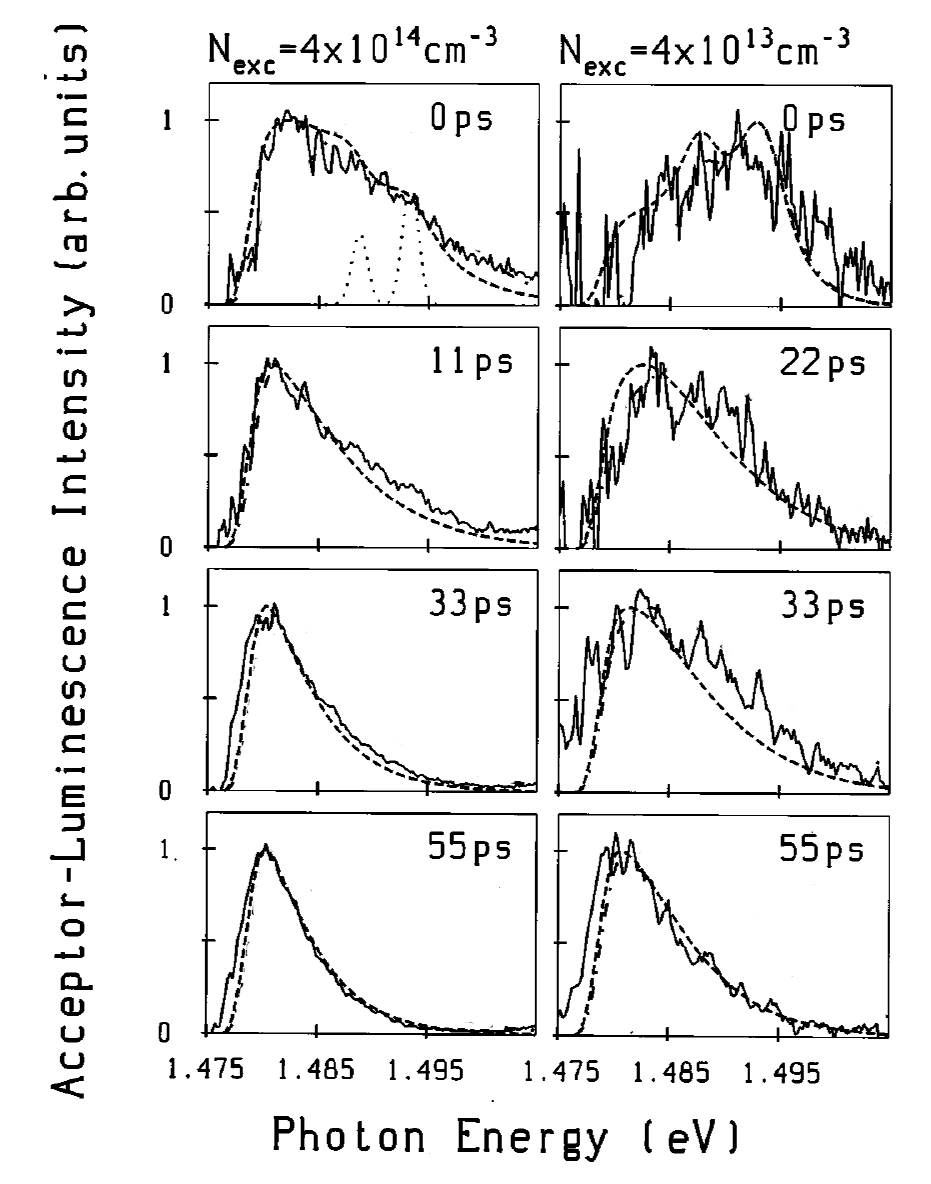}
\end{center}
\caption{Time evolution of the free electron energy distribution in bulk GaAs following their generation by a short (2 ps) laser pulse, for two different densities. Solid lines are the data, deduced from the electron-acceptor luminescence, and dashed lines are the quantum Boltzmann equation theory, including carrier-carrier, carrier-phonon, and carrier-impurity scattering. The dotted line in the upper left panel is the predicted initial population due to the laser excitation.
From Ref.~\protect\cite{graz-ee}.}
\label{ee}
\end{figure}

The same method can be applied to free electrons and holes in semiconductors. Figure \ref{ee} shows a measurement of the time-resolved energy distribution of free electrons in the semiconductor GaAs, also recorded in the laboratory of Wolfgang Ruehle in Stuttgart \cite{gaas}. The energy distribution of the electron gas is seen in the spectrum of the electron-acceptor luminescence. As seen in this figure, at low density the electron energy distribution evolves from a nonequilibrium to an equilibrium distribution in about 50 ps.  The data could be reasonably well fit with quantum Boltzmann equation model \cite{graz-ee} which included carrier-carrier scattering, carrier-phonon scattering, and carrier-impurity scattering (the dopant atoms in the $p$-doped GaAs used for these experiments play a major role in the thermalization). 
\begin{figure}[<float>]
\sidecaption
\includegraphics[width=0.43\textwidth]{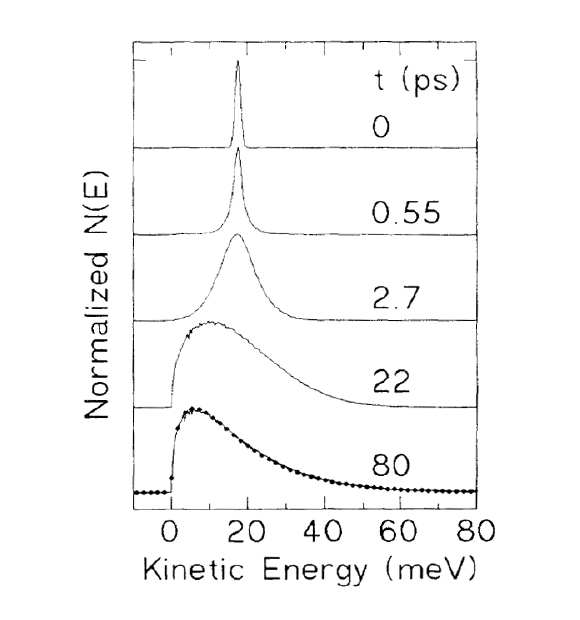}
\caption{Time evolution of the free electron energy distribution in for the parameters of bulk GaAs, computed by solution of the quantum Boltzmann equation, for constant electron screening. The dots in the last curve are a fit to a Maxwell-Boltzmann distribution. From Ref.~\protect\cite{gaas}.}
\label{eeMB}
\end{figure}

The modeling of the electron-electron (and electron-hole and hole-hole) interaction is much more complicated than the modeling of exciton-exciton interactions, which can be treated as short-range, hard-sphere interactions. The interaction potential for electrons  is (see Ref.~\cite{snokebook}, Section 8.10)
\begin{equation}
U(k_1,k_2,k_3,k_4) = \frac{e^2/\epsilon}{|k_1-k_3|^2 + \kappa^2},
\label{coul}
\end{equation}
where again, momentum conservation is assumed, which gives $\Delta \vec{k} = \vec{k}_1-\vec{k}_3 = \vec{k}_4 - \vec{k}_2$. The parameter $\kappa$ is the screening parameter. Without this parameter, the Coulomb interaction diverges at $\Delta k = 0$. In the long wavelength, low energy approximation, the screening is given by
\begin{equation}
\kappa^2 = -\frac{1}{V} \frac{e^2}{\epsilon}\int  \frac{\partial n(E)}{\partial E}D(E) dE
\end{equation}
in a three-dimensional system. This formula gives the familiar Debye and Thomas-Fermi screening results for equilibrium distributions. For nonequilibrium distributions, the screening can be calculated for the distribution $n(E)$ at each point in time \cite{eedens} as the distribution is evolved by iteration. Similar formulas can be deduced for a two-dimensional system \cite{eedens}. 

The general theory of nonequilibrium electron-hole scattering, and plasma scattering in general, is still an open field of theory.  The above formulas should work well on long time scales, but on very short time scales, they will break down, under the same conditions when Fermi's golden rule will break down, namely, when the rate of change of the energy distribution is so fast that the equivalent energy $\Delta E \sim \hbar/ \Delta t$ is larger than the range over which the energy distribution and scattering matrix element can be considered constant (see Ref.~\cite{snokebook}, Section 4.7.) The rapid variation of the Coulomb scattering interaction makes this energy range very narrow. In this limit, a method must  be developed to generalize the quantum Boltzmann equation to take the phase of the electrons into account, known as a quantum memory kernel  \cite{haugqmk,yanqmk,mathqmk,chemqmk,preqmk}.  Various methods of solving the quantum Boltzmann equation for electron-electron and electron-hole scattering were presented in the 1990's \cite{eedens,graz-ee,french,haug,binder}, but none of them could provide fits to time-resolved data as accurate as that shown in Fig.~\ref{cu2o}, although the fit shown in Fig.~\ref{ee}, from Ref.~\cite{graz-ee}, is not bad. A model for plasma scattering should properly include electron-electron, electron-hole, and hole-hole scattering, with exchange, as well as interaction of electrons and holes with phonons and elastic scattering with impurities. These latter processes are easy to model, but the Coulomb scattering of the carriers with exchange is much harder to calculate numerically. The exchange process, in particular, which can be dominant, depends on $\vec{k}_1-\vec{k}_4$, which cannot be written simply in terms of $\Delta \vec{k} = \vec{k}_1 - \vec{k}_3$ and $k_{cm} = \frac{1}{2}(\vec{k}_1+\vec{k}_2)$. The methods discussed above which make the calculation quite simple for the case of an isotropic, homogeneous gas therefore cannot be used. This does not make the quantum Boltzmann equation inapplicable, it just means that a Coulomb scattering problem requires much more computational time. 

The quantum Boltzmann equation allows us to make some interesting conclusions about the power laws for various types of processes in various density regimes. Because the Coulomb scattering interaction (\ref{coul}) is so sharply peaked around $\Delta k = 0$, the rates deduced in different experiments for effects that depend on carrier-carrier scattering in plasmas can vary enormously. Ref.~\cite{eedens} calculated the intrinsic power laws for three different types of experiments. One type can be called {\em dephasing} measurements, which depend on the total carrier-carrier scattering rate. This type includes most four-wave mixing experiments. Another can be called {\em momentum relaxation} experiments. This type includes most transport experiments. The last can be called {\em energy relaxation} measurements. This includes experiments like that shown in Fig.~\ref{eeMB} which measure the time scale for thermalization to an equilibrium energy distribution. Using a self-consistent model for dynamic screening, but neglecting exchange of identical particles, Ref.~\cite{eedens} deduced that the intrinsic dephasing rate will be independent of density, that the momentum relaxation will be proportional to density, and the energy relaxation rate will be proportional to the square root of density, in three dimensions. The reason why the rate for total electron-electron scattering is constant as density falls is that the screening becomes less efficient at low densities, so that the scattering cross section for electron-electron scattering increases in just such a way as to keep the total scattering constant as the density decreases. However, the scattering will more and more favor forward scattering, that is, scattering with little or no change in an electron's momentum and energy, as the density decreases. Thus the momentum and energy relaxation rates will fall. 

The surprising prediction of constant dephasing rate independent of density \cite{eedens} was approximately confirmed by four-wave mixing experiments done in the 1990's which showed very weak dependence of the dephasing rate on density. One study \cite{arlt} found that the dephasing rate increased by less than 40\% over a factor of ten increase of density from $10^{16} - 10^{17}$ cm$^{-3}$, corresponding to a power law of $1/\tau \propto n^{0.15}$; another study \cite{wegener} found about 50\% increase over the same range. The absolute values of the dephasing in these studies were also consistent with the quantum Boltzmann theory. For the limit of low density (i.e., when Fermi statistical state filling is negligible), the dephasing rate deduced from the quantum Boltzmann equation, neglecting exchange, is (for the derivation see \cite{snokebook}, Section 5.5)
\begin{equation}
\frac{1}{\tau} = \frac{\sqrt{2}e^2(mk_BT)^{1/2}}{\pi^{3/2}\epsilon\hbar^2},
\label{dephase}
\end{equation}
which depends only on the mass and temperature, i.e. the average thermal velocity of the particles. This formula gives $\tau \simeq 25$ fs for electrons in GaAs at room temperature and $\tau \simeq 200$ fs at for electrons in GaAs at 4 Kelvin. Ref.~\cite{arlt} found an average low-density dephasing time of around 15 fs at room temperature, while Ref.~\cite{wegener} found an average dephasing time of around 25 fs. Both are much shorter than the electron-phonon scattering time of around 150 fs \cite{phon1,phon2}.

One of the most controversial results of the work on plasma scattering in the 1990's was the claim of the group of Shank \cite{shank1,shank2} that in the high density regime, the electron dephasing rate scales as $n^{1/3}$ in three dimensions and as $n^{1/2}$ in two dimensions. These power laws could be justified by a quasi-phenomelogical estimate of the dependence of the screening length on density \cite{eedens}. 
Later experimental work \cite{wegener}, however, indicated that the power laws reported by Shank and coworkers might not apply across a very wide range of density. 

Several experimental factors made these experiments difficult to interpret. One issue is that ultrafast laser pulses have broad spectral width, and part of the laser spectrum can overlap with the excitonic range of the spectrum, giving interference effects. This makes experiments at very low temperature, where exciton effects are important, hard to interpret. Another issue is that the higher power laser pulses used to create higher carrier density can also lead to higher temperature. Equation (\ref{dephase}) predicts that the dephasing rate should increase as $T^{1/2}$. If the temperature increases weakly with increasing density, this could explain the non-constant dephasing rate seen in the experiments. The difference in the absolute values of the dephasing times seen in Refs.~\cite{arlt} and \cite{wegener} can also be understood as due to different effective temperatures of the electrons in different excitation conditions, which could include different pump laser photon energies and different rates of heat flow out of the samples. 

In general, there are still many open questions in the topic of the intrinsic power laws of carrier-carrier scattering, and a need for clear experiments and theoretical models that take into account dynamical screening and exchange. Kira and Koch \cite{kochkira} have led the way in numerical models of carrier scattering, but their numerical methods are too complicated for many others to reproduce, and there is a need for simpler analytical approaches that work well in the middle ground between the manifestly incorrect relaxation time approximation and a full quantum wave function model.  The quantum Bolztmann results clearly show that we cannot talk meaningfully of ``a'' rate for carrier-carrier scattering; this is in sharp contrast to the relaxation-time approximation which is often used in the theory of carrier scattering, which writes a single characteristic relaxation time for each scattering process.

\section{Results for Polaritons}

\begin{figure}[<float>]
\sidecaption
\includegraphics[width=0.45\textwidth]{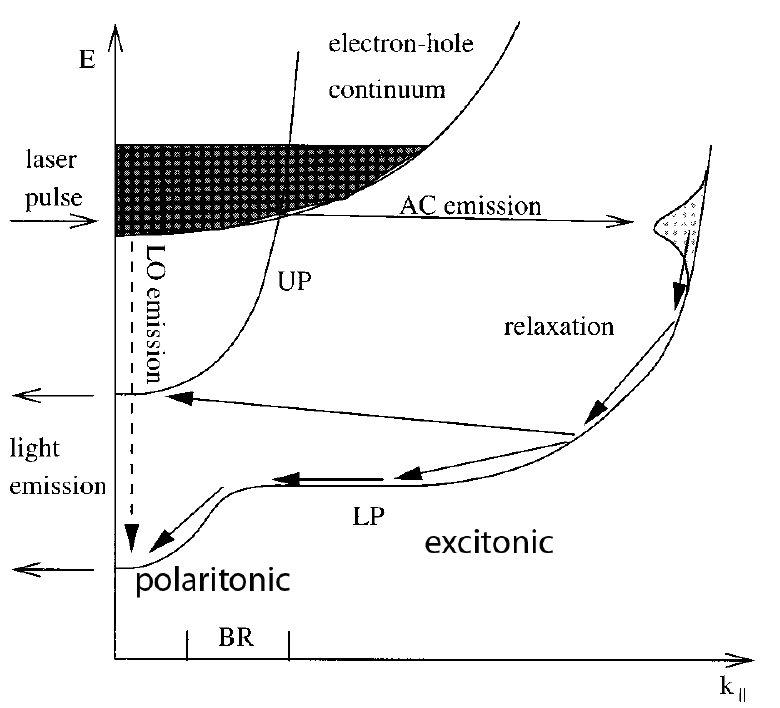}
\caption{Schematic illustration of the bands and relaxation processes in the polariton generation process. The lowest exciton band is split by the cavity into a lower polariton band (LP) and upper polariton band (UP). The lower polariton band goes from the polariton minimum at zone center  to an excitonic region, with a connecting bottleneck region in between (labeled BR). The polaritons have an effective mass much lighter than that of the excitons. From Ref.~\protect\cite{tassone}.}
\label{polschem}
\end{figure}

In the past ten years, microcavity exciton-polaritons have become a major field of study in optics, because they can be viewed as a Bose-Einstein condensate (BEC) of light-mass, weakly-interacting particles \cite{pt}, with many fascinating effects such as vortices \cite{vort1,vort2}.  Exciton-polaritons are superpositions of photons and excitons in a semiconductor; the modern experiments use a microcavity, fabricated using two Bragg mirrors made of alternating dielectrics, to produce cavity photon states which are resonant with the exciton states in quantum wells inside the cavity. (For reviews, see Refs.~\cite{kav} and \cite{PT}.) This produces poalritons in a two-dimensional plane with very light effective mass, around $10^{-4}$ times the vacuum electron mass, which interact with each other and with phonons due to their excitonic component. 
The light mass of the cavity polaritons allows the possibility of a BEC at room temperature, while the interactions between the polaritons lead to many useful nonlinear effects \cite{kav}. Figure \ref{polschem} shows the basic band structure of the microcavity polaritons and the scattering processes following generation of free electrons and holes. 

\begin{figure}[<float>]
\sidecaption
\includegraphics[width=0.63\textwidth]{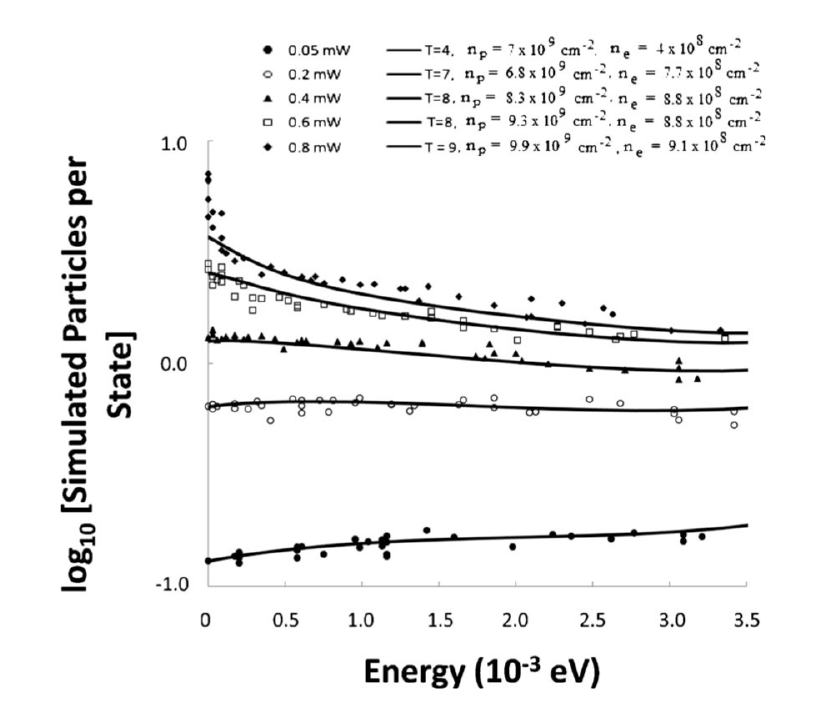}
\caption{Symbols: the energy distribution of a polariton gas in steady state, for several generation rates. Solid lines: fit to the data using a quantum Boltzmann equation solution. From Ref.~\protect\cite{hartwell}.}
\label{pol}
\end{figure}

A quantum Boltzmann model is essential for understanding the energy distribution of the polariton gas. A polariton in this type of system only lives for around 10-20 picoseconds before turning into a photon external to the cavity which exits irreversibly.  A typical microcavity polariton condensate experiment therefore involves a quasi-cw pump which creates hot carriers that eventually become hot polaritons, which collide with each other and emit phonons to settle down and thermalize in the low energy states. Figure~\ref{pol} shows the experimentally measured energy distribution of a polariton gas for several densities below and near the BEC threshold. Well below the threshold, the gas is not thermalized at all; as it approaches the threshold it becomes more thermalized, and right near the BEC threshold a peak appears in the energy distribution near the ground state. 

Numerous models \cite{tassone,haug2,malp} using a quantum Boltzmann approach have been used to model microcavity exciton-polariton behavior; this work is surveyed in Ref.~\cite{hartwell}. The general conclusion of all this work is that it is indeed proper to speak of the polariton gas as a condensate, in which the Bose statistics of the polaritons plays the key role in the buildup of the population in low-energy states.  

The full energy distribution, which includes not just the polariton states, but also the exciton states at higher energy and the bottleneck region which connects the two (see Fig.~\ref{polschem}), is never well described by an equilibrium distribution with single temperature, because the coupling of the exciton population to the polariton population through the bottleneck region is quite weak. However, a quantum Boltzmann fit to the steady-state energy distribution, shown in Fig.~\ref{pol}, shows that the polaritons are nearly equilibrated among themselves. These fits were obtained by evolving a nonequilibrium polariton gas in time until it reached steady state.  The model shows that the collisions between the polaritons play the key role in the buildup of polaritons in the low-energy states. Their interaction with acoustic phonons is so weak that the gas can never thermalize through phonon interactions alone.

The Boltzmann model used to fit the data in Fig.~\ref{pol} does not include any coherent phase information, and therefore cannot model the condensate itself. Laussy and coworkers \cite{laussy-coh} and Haug and coworkers \cite{haug-coh} have used modifications of the quantum Boltzmann equation to account for the coherence of polaritons, as will be discussed in the next section. 

A surprising experimental result of the polariton experiments has been that at very low polariton density, the polaritons still thermalize reasonably well, much faster than expected from polariton-phonon interaction, although the polaritons do not reach equilibrium at low density. Ref.~\cite{hartwell} showed that this low density quasi-equilibrium can be explained as due to the interaction of polaritons with free electrons, which can exist as a permanent population with low density in the solid. Polariton-electron interaction has been shown to be efficient \cite{mal-ee}. Another approach to the low density behavior was taken in Ref.~\cite{mal-sat}, which treated elastic scattering of polaritons with disorder by replacing the energy-conserving delta-function in the quantum Boltzmann equation with a Lorentzian, allowing violation of energy conservation. While this approach might seem natural, since scattering can lead to energy broadening of transition rates in Fermi's golden rule (see Section 8.4 of Ref.~\cite{snokebook}), it is problematic in the case of the many-body quantum Boltzmann equation. Work by Haug and coworkers in the early 1990's showed that relaxing strict energy conservation in the Boltzmann equation leads to an overall instability of the total energy of an interacting gas, when the particles have infinite lifetime. In the case of polaritons with finite lifetime, this instability is avoided, but replacing the energy-conserving delta-function with a Lorentzian in the quantum Boltzmann equation has not been well theoretically justified. 

\section{Modifications for Condensates}

In general, the question of the onset of coherence in a condensate, starting from a system which is initially incoherent, has been an active area of theory in the past twenty years. The issue first became important for the theory of condensates of excitons \cite{snokewolfe}, because excitons have finite lifetime, and therefore there were questions whether they could undergo condensation within their lifetime. The topic was next picked up for atomic condensates in traps \cite{bec1,bec2,bec3,gard}, because the atoms in these traps also have finite lifetime; in this case their lifetime is not determined by recombination, but rather their rate of escaping the trap. In recent years the issue has come to prominence again because of the successes of polariton condensates, discussed above. 

\begin{figure}[<float>]
\sidecaption
\includegraphics[width=0.6\textwidth]{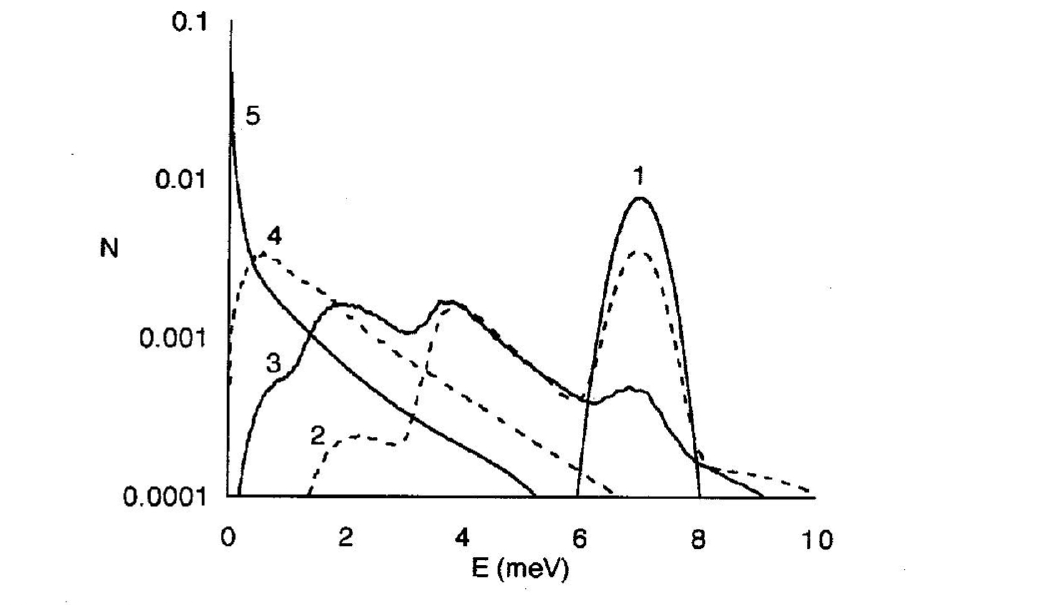}
\caption{Quantum Boltzmann equation solution of a gas of excitons, with the same properties as those in Fig.~\protect\ref{cu2o}, when the exciton density is above the critical threshold for Bose-Einstein condensation, and the time evolution is entirely due to interaction with acoustic phonons. Lattice temperature was 16 K; exciton density $7\times10^{18}$ cm$^{-3}$; the curves labeled 1-5 (alternating solid and dashed lines) correspond to times after the initial creation of  0, 11, 42, 120, and 832 picoseconds, respectively. From Ref.~\protect\cite{moskbook}.}
\label{phoncond}
\end{figure}

As discussed above, the quantum Boltzmann equation does not directly model condensation, because it does not account for the phase of the particles, and true condensation consists of spontaneous phase coherence. However, a quantum Boltzmann equation can account for the buildup of the macroscopic population in the ground state. The early work modeling excitons \cite{snokewolfe} showed that a two-body, short-range interaction between bosons led to exponentially increasing population of the ground state, i.e. $n_0 \sim e^{t/\tau}$, where $\tau$ is of the order of the classical scattering time defined by $1/\tau = n\sigma v$, where $n$ is the particle density, $\sigma$ the scattering cross section, and $v$ the average velocity.  Later work \cite{russian} showed that the motion of the population toward the ground state via two-body, short-range scattering follows an interesting power law. The ground state can also obtain a macroscopic population entirely through phonon emission and absorption, even when there is no interparticle scattering. Figure~\ref{phoncond} shows the theoretical time evolution of a gas of excitons with the same properties as those in Fig.~\ref{cu2o}, with density high enough to lead to Bose statistical effects (accounted for through the $(1+n_f)$ factors) but with the exciton-exciton two-body interaction artificially turned off. The gas of excitons quickly acquires a large population in the ground state above the critical density for Bose condensation.  

It turns out that for a polariton gas, the interaction with phonons is not sufficiently fast to bring about condensation, because of both the short lifetime of the polaritons and the ``bottleneck'' in the density of states of the polaritons, which makes it difficult for excitons in high energy states to scatter into polariton states. The two-body scattering between the polaritons therefore plays the dominant role in the buildup of the condensate \cite{hartwell}. 

Laussy, Malpuech, Kavokin, and coworkers \cite{laussy-coh}, building on earlier work by Gardiner and Zoller and coworkers \cite{gard},
showed that the coherence of the condensate can be accounted for by generalizing the quantum Boltzmann equation to a similar equation for the density matrix $\rho = |\psi \rangle\langle \psi|$. For the case of excitons or polaritons interacting with phonons, they found, for the ground state, 
\begin{equation}
\frac{\partial \rho_0}{\partial t} = \frac{1}{2}W_{\rm in}(t)(2a_0^{\dagger}\rho_0a_0 -a_0a^{\dagger}_0 \rho_0 -\rho_0 a_0a^{\dagger}_0)
+ \frac{1}{2}(\Gamma_0+W_{\rm out}(t))(2a_0\rho_0a_0^{\dagger} -a_0a^{\dagger}_0 \rho_0 -\rho a^{\dagger}_0a_0),
\end{equation}
where $W_{\rm in}(t)$ and $W_{\rm out}(t)$ are the in-scattering and out-scattering rates for the ground state from the quantum Boltzmann equation, and $\Gamma_0$ is the recombination rate of the ground state. 
From this they obtained an equation for the expectation value of the amplitude,
\begin{equation}
\frac{\partial}{\partial t}\langle \psi| a | \psi \rangle = (W_{\rm in}(t)-W_{\rm out}(t))\langle \psi| a | \psi \rangle.
\label{a}
\end{equation}
Equation (\ref{a}) has the same form as the equation for onset of lasing (for a review see Ref.~\cite{snokebook}, Section 11.3).  This form of equation, for both condensates and lasers, shows that a small coherent amplitude will be amplified and grow exponentially. If the initial amplitude is strictly zero, of course there will be no growth of the coherent amplitude. The coherent amplitude $\langle a \rangle$ must therefore be given a ``seed'' in the numerics which will be amplified. This is a direct expression of the very basic effect of spontaneous symmetry breaking which occurs in condensates. Nothing in the equations chooses the exact phase of the condensate; this must come from some external seed, or fluctuation. 

This approach has the limitation that one state, the ground state, is ``cut out'' from the rest of the quantum states somewhat artificially, but this is in the spirit of the way that Einstein treated the condensate state separately from others. Another approach is to treat the full continuum of states near $k=0$, in what is called a ``classical field'' approach \cite{class1,class2,class3}, adapted from quantum optics of lasers.  This approach treats all the states near the ground state, but has the disadvantage that it does not transition continuously from these states to the thermal states well above the ground state.

A few words are in order about the role of the seed in the generalized Boltzmann models like that of Ref.~\cite{laussy-coh}. No seed was used in the iterated solution shown in Fig.~\ref{phoncond}, nor in the two-body scattering model of Ref.~\cite{snokewolfe}.  Thus the growth of the {\em population} in the ground state is not dependent on the symmetry breaking. However, the {\em coherence} of the ground state does depend on a seed which breaks the symmetry. 

In the model of Fig.~\ref{phoncond}, as well as the two-body model of Fig.~\cite{snokewolfe} and similar models, the energy distribution of the particles is represented by a set of values of $n(E_i)dE_i$ for a set of energy ranges $(E_i,E_i+dE_i)$. The ``condensate'' population in these models is represented by the value of $n(0)$ for the lowest energy range $(0,dE_0)$.  Strictly speaking, this method therefore cannot distinguish between a ``true'' condensate in the exact $E=0$ state and a ``quasicondensate'' in states very near $E=0$. However, what one can do in a quantum Boltzmann approach is to use a grid of ever smaller and smaller energy intervals near $E=0$, and show that the lowest of these energy intervals is always the one which eventually jumps up in population, no matter how small the energy range near the ground state. 

In the two-dimensional polariton system, the dynamics of condensation is not much different from the three-dimensional models. Technically, a true condensate cannot appear in a translationally invariant two-dimensional system \cite{kt,berman2d}.  However, in a finite system, the coherence length can be long compared to the size of the system, so that the lack of long-range order predicted for a two-dimensional system does not play a role. 

\section{Conclusions}

The quantum Boltzmann equation, or Fokker-Planck equation, has a wide range of applicability to nonequilibrium dynamics, and many spectroscopy experiments provide data which can be directly compared to solutions of the quantum Boltzmann equation. The solutions for the quantum Boltzmann equation for excitons interacting either with phonons or with each other via short-range interactions are well worked out, and in the case of exciton-phonon interaction there is outstanding agreement of theory and time-resolved experiments. The status of the theory and experiments for the evolution of a plasma is much less settled. However, the results of the quantum Boltzmann approach for a gas interacting via Coulomb potential at low density give power laws and absolute scattering rates that are basically consistent with the experiments, including a prediction \cite{eedens} of constant dephasing rate at low density and reasonable fits to time-resolved data \cite{graz-ee}. 

The quantum Boltzmann approach also describes the behavior of Bose-Einstein condensates.  The Bose statistics give rise to a macroscopic occupation of the ground state in a system which starts far from equilibrium, with no condensate, both for systems interacting with a phonon bath and for closed systems interacting via number-conserving two-body scattering. Quantitative fits to data have been performed for polariton condensate systems.  The quantum Boltzmann equation can also be coupled to a general density matrix approach for the evolution of the phase of a Bose condensate. 

Thus, despite its simplicity compared to full quantum Monto Carlo and quantum memory kernel approaches, the quantum Boltzmann equation has wide applicability to quantitative experiments measuring the particle distribution of gases, in particular as measured by optical spectroscopy of semiconductors.

\begin{acknowledgement} 
This work has been supported by the National Science Foundation under grant DMR-0706331. 
\end{acknowledgement}

\end{document}